\documentclass[a4paper]{jpconf}
\usepackage{graphicx}

\def\apjl{Astrophys.\ J.\ Lett.\ }

\def\mnras{Mon.\ Not.\ Roy.\ Astron.\ Soc.\ }

\def\prd{Phys.\ Rev.\ D\ }

\def\jcap{JCAP}
\usepackage{citesort}
\begin{document}
\vspace*{-0.1cm}
\title{Constraints on asymmetric dark matter from asteroseismology}

\author{Jordi Casanellas$^{1,2}$, Il\'idio Lopes$^{2,3}$}

\address{$^1$ Centro de Astronomia e Astrof\'isica da Universidade de Lisboa, Portugal}
\address{$^2$ CENTRA, Instituto Superior T\'ecnico, Universidade T\'ecnica de Lisboa, Portugal}
\address{$^3$ Departamento de F\'isica, Universidade de \'Evora, Portugal}

\ead{jordicasanellas@ist.utl.pt ; ilidio.lopes@ist.utl.pt}

\begin{abstract}
We report recent results on the impact of asymmetric dark matter (DM) particles on low-mass stars. First, we found that the small convective core expected in stars with masses between 1.1 and 1.3 M$_{\odot}$ is suppressed due to DM cooling. Moreover, stars with masses below 1~M$_{\odot}$ have their central temperatures and densities more strongly influenced by DM than in the solar case. We were able to put limits to the DM mass and spin-dependent DM-proton scattering cross section by comparing the modelling of the nearby star $\alpha$~Cen B with photometric, spectroscopic and asteroseismic observations.
\end{abstract}

\section{Introduction}
If the dark matter (DM) of the Universe is made of weakly interacting massive particles (WIMPs) their accumulation inside stars may lead to observable effects that depend on the unknown characteristics of the dark particles~\cite{1985ApJ...294..663S,Lopes:2002gp,Spolyar:2007qv,Iocco:2008xb,Scott:2008ns,Casanellas:2009dp,Gondolo:2010kq,Zackrisson:2010jd,Sivertsson:2010zm,Casanellas:2010he,Scott:2011ni,Casanellas:2011qh,Zentner:2011wx,2012MNRAS.422.2164I,Li:2012qf,2013PhRvD..87l3506L}. Thus, the DM parameter space can be constrained using this approach, in particular in the case of models of DM particles that do not annihilate in stellar interiors. A good example is asymmetric DM \cite{1987NuPhB.283..681G,2008PhRvD..78f5040K,Davoudiasl:2011fj,2012NJPh...14i5011D,Cirelli:2011ac,Tulin:2012re,Blennow:2012de,MarchRussell:2012hi,Gu:2012fg,Pearce:2013ola,2013arXiv1306.5878B,McCullough:2013jma}, a DM model motivated to explain the similar cosmic abundance of dark and baryonic matter and that naturally predicts a DM 
mass of the 
order of the few GeV (see for instance the recent reviews~\cite{Petraki:2013wwa,Zurek:2013wia}).\\

A fraction of the WIMPs in the galactic halo may scatter with nucleons of the stellar plasma and become gravitationally captured by stars~\cite{art-Gould1987,art-GondoloEdsjoDarkSusy2004,Lopes:2011rx}. The efficiency of this process increases with the stellar density (among other factors), so compact stars can be in principle strongly influenced~\cite{Moskalenko:2007ak,Bertone:2007ae,McCullough:2010ai,2011PhRvD..84j3510F,Bramante:2013hn}. The destruction of neutron stars by black hole formation in their interior has been predicted, and limits on the properties of asymmetric DM have been derived using this approach~\cite{Kouvaris:2010jy,Kouvaris:2011fi,Goldman:2013qla,2013PhRvD..87l3537K,2013PhRvD..87l3507B}.\\

In the case of the Sun and solar-type stars, the conduction of DM particles in the stellar interior acts as a new energy transport mechanism, cooling the center of the stars~\cite{Bottino:2002pd,art-Taosoetal2010PhRvD,Cumberbatch:2010hh,Lopes:2012af,2012ApJ...746L..12T,2012ApJ...752..129L,Iocco:2012wk}. The impact on the properties of these stars tends to be less significant than for compact stars, but on the other hand the physics of the interior of solar-type stars is much better known and observations are much more precise, in particular with the complementary diagnostic of the solar interior provided by solar neutrino measurements~\cite{2010Sci...330..462L} and helioseismology~\cite{Lopes:2010fx}.\\

In these proceedings we report the study of the impact of asymmetric DM on stars with masses similar to that of the Sun~\cite{Casanellas:2012jp}, stars embedded in halos of DM with a density equal to that estimated for the solar neighbourhood, $\rho_\chi=0.4\;$GeV cm$^{-3}$~\cite{Garbari:2012ff}. We focus on stars whose acoustic oscillations have already been precisely identified. The characteristics of these stars are shown in Table~\ref{tab-starchars}. Our approach is motivated mainly by the following reasons: i) stars with lower masses are more strongly influenced by DM, ii) stars with a small convective core (1.1-1.3~M$_{\odot}$ stars) can have their dominant energy transport mechanisms altered due to DM, and iii) asteroseismology (the study of the stellar oscillations) is a technique sensitive to properties of the stellar cores, where the DM impact occurs.\\

As we will show, we were able to put the first limits to the nature of DM using an asteroseismic analysis of $\alpha$ Cen B. The reader is addressed to reference~\cite{Casanellas:2012jp} for a more thorough description of the methods used and the results obtained.\\

\begin{table}
\caption{Summary of observational constraints on the modelling and selected results.\label{tab-starchars}}
\begin{center}
\begin{footnotesize}
\begin{tabular}{l r r r r r r}
 \br
Star & $M$ (M$_{\odot}$) & $L$ (L$_{\odot}$) & $T_{eff}$ (K) & $(Z/X)_s$ & $\langle\Delta \nu_{n,0}\rangle ^{a}$ ($\mu$Hz) & $\langle\delta \nu_{02}\rangle ^{a}$ ($\mathit{\mu}$Hz)\\
\mr
\multicolumn{2}{l}{\textbf{KIC 8006161}} &&&&& \\
\hspace{0.1cm}Observations $^{b}$ & 0.92-1.10 & $0.61 \pm 0.02$ & $5340 \pm 70$ & $0.043 \pm 0.007$ & $148.94 \pm 0.13$& $10.10 \pm 0.16$ \\
\hspace{0.1cm}Stand. model. & 0.92 & 0.63 & 5379 & 0.039 & 149.03  & 10.12 \\
\hspace{0.1cm}DM model.$^{c}$ & 0.92 & 0.63 & 5379 & 0.039 & 149.08  & 9.13 \\
\multicolumn{2}{l}{\textbf{HD 52265}} &&&&& \\
\hspace{0.1cm}Observations $^{b}$ & 1.18-1.25 & $2.09 \pm 0.24$ & $6100 \pm 60$ & $0.028 \pm 0.003$ & $98.07 \pm 0.19$ & $8.18 \pm 0.28$ \\
\hspace{0.1cm}Stand. model.& 1.18 & 2.22 & 6170 & 0.028 & 97.92 & 8.16 \\
\hspace{0.1cm}DM model.$^{c}$ & 1.18 & 2.22 & 6170 & 0.028 & 98.05 & 7.65 \\
\multicolumn{2}{l}{\textbf{$\alpha$ Cen B}} &&&&& \\
\hspace{0.1cm}Observations $^{b}$ & $0.934 \pm 0.006$ & $0.50 \pm 0.02$ & $5260 \pm 50$ & $0.032 \pm 0.002$ & $161.85 \pm 0.74$ & $10.94 \pm 0.84$ \\
\hspace{0.1cm}Stand. model.& 0.934 & 0.51 & 5245 & 0.031 & 162.56 & 10.23 \\
\hspace{0.1cm}DM model.$^{c}$ & 0.934 & 0.51 & 5230 & 0.031 & 162.45 & 8.95 \\
\br
\end{tabular}
$^{a}$ The averages are calculated for the intervals $2750<\nu$($\mu$Hz)$<3900$ for KIC 8006161, $1600<\nu$($\mu$Hz)$<2600$ for HD 52265, and $3300<\nu$($\mu$Hz)$<5500$ for $\alpha$~Cen~B.\\
$^{b}$Data from \cite{Mathur:2012sk} and \cite{Bruntt:2012hs} for KIC 8006161, \cite{Ballot:2011kn} for HD 52265, and \cite{Kjeldsen:2005td} for $\alpha$~Cen~B.\\
$^{c}$ $m_{\chi}=5\;$GeV, $\sigma_{\chi,SD}=3\cdot10^{-36}\;$cm$^2$, $\rho_{\chi} = 0.4\;$GeV cm$^{-3}$.
\end{footnotesize}
\end{center}
\end{table}

\section{Results}

\subsection{Impact on the central temperatures and densities. Suppression of core convection}
We modelled the star KIC 8006161, reproducing all its observed characteristics, including the large and small frequency separations as measured by the \textit{Kepler} mission. We found a $\sim10\%$ decline in its central temperature when the influence of the accumulation of asymmetric DM particles with $m_{\chi}=5\;$GeV and $\sigma_{\chi,SD}=3\cdot10^{-36}\;$cm$^2$ was taken into account.\\

Furthermore, we also studied the star HD 52265, a 1.2 M$_{\odot}$ star whose oscillation frequencies have been measured by the \textit{CoRoT} mission. We found that, whereas in the standard modelling this star is expected to have a convective core, the inclusion of the DM energy transport removes this core and the star is left with a radiative interior. Specific combinations of frequencies of low-degree modes have been shown to be sensitive to the presence of small convective cores (see Ref.~\cite{2011A&A...529A..10C}). We leave this interesting diagnostic approach for a further work.\\

\subsection{Constraints on the DM properties}
We have chosen to study the star $\alpha$~Cen B because it is close to the Sun and belongs to a binary system, which has allowed a very accurate measurement of its mass and other characteristics (see Table~\ref{tab-starchars}). This fact strongly reduces the uncertainties in the modelling of the star. We used the stellar oscillations, in particular the mean small separation between the modes of low degree $\langle\delta \nu_{02}\rangle$, to identify the modifications introduced by asymmetric DM in the core of the star.\\

\begin{figure}[t]
 \centering
 \includegraphics[scale=1.4]{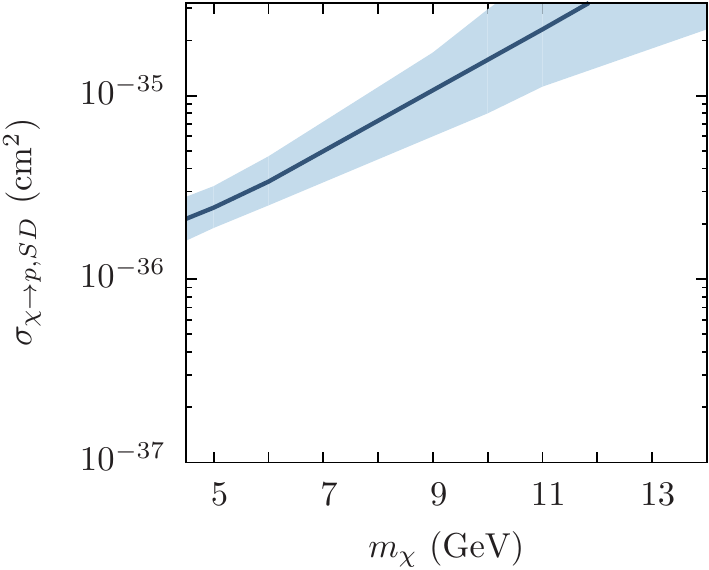}
 \caption{Upper limits for the WIMP-proton spin-dependent scattering cross section as a function of the WIMP mass from an asteroseismic analysis of the star $\alpha$~Cen B. Asymmetric DM particles with properties above the blue line produce a strong impact on the core of the star, leading to a mean small frequency separation more than 2~$\sigma$ away from the observations. The filled region shows the uncertainty in the modelling when the observational errors are taken into account. A density of $\rho_{\chi}=0.4\;$GeV cm$^{-3}$ was assumed. Figure adapted from Ref.~\cite{Casanellas:2012jp}.}
\label{fig-ADMconstraints}
\end{figure}

We found that the stellar models strongly influenced by the DM cooling mechanism, while reproducing all the global properties of $\alpha$~Cen~B within observational errors, cannot account for the observed small frequency separations. The existence of asymmetric DM particles with $m_{\chi}$ and $\sigma_{\chi,SD}$ above the blue line in Figure~\ref{fig-ADMconstraints} can be excluded because they lead to models of $\alpha$~Cen~B with a $\langle\delta \nu_{02}\rangle$ more than 2~$\sigma$ away from the observed value. The 1~$\sigma$ variation on $\langle\delta \nu_{02}\rangle$ when the classical stellar parameters ($M$, $L$, $T_{eff}$ and $(Z/X)_s$) vary within the observational errors is plotted as a filled region around the blue line. Further uncertainties not included in this analysis are those in other parameters of stellar modelling, such as nuclear reaction rates and chemical abundances, with an expected minor impact on our diagnostic and, more importantly, the error in the determination of the local 
DM density, which will introduce an estimated uncertainty of a factor smaller than 2 in $\sigma_{\chi,SD}$~\cite{Garbari:2012ff,Pato:2010yq}.\\ 

Our method provides a competitive exclusion plot on $\sigma_{\chi,SD}$ for low-mass DM particles. These constraints are particularly valuable taking into account the present controversy among different direct detection experiments in the same region of the DM parameter space~\cite{Buckley:2013gjo}.

\par\vfill\break 

\advance\vsize by 1cm 
\advance\voffset by -1cm 

\section*{References}
\providecommand{\newblock}{}

\par\vfill\break 

\advance\vsize by -0.6cm 
\advance\voffset by 0.6cm 
\end{document}